\documentstyle[aps,epsf,preprint]{revtex}
\def\myfigure#1#2{{\leftskip=0.10753\textwidth \rightskip\leftskip\small
\begin{figure}\baselineskip=14pt plus 2pt minus 1pt
\centerline{#1}\nobreak\smallskip\nobreak #2\end{figure}}}
\global\firstfigfalse
\begin{document}
\draft
\preprint{IMSc./2000/08/50}
\title{Numerical evidence for monopoles in 3-dimensional Yang-Mills theory}
\author{Pushan Majumdar\footnote{Present address: Tata Institute of Fundamental 
Research, Homi Bhabha Road, \\Mumbai 400005, India.} 
\thanks{e-mail:pushan@imsc.ernet.in} \and
\addtocounter{footnote}{1}
Dong-Shin Shin \thanks{e-mail:shin@imsc.ernet.in}}
\address{Institute of Mathematical Sciences, C.I.T. Campus, Taramani,
Madras 600-113.}
\maketitle
\begin{abstract}
Recently Anishetty, Majumdar and Sharatchandra have proposed a way of characterizing 
topologically non-trivial configurations for 2+1-dimensional Yang-Mills theory in a 
local and manifestly gauge invariant manner. In this paper we develop criteria to locate such
objects in lattice gauge theory and find them in numerical simulations.  
\end{abstract}
\section{Introduction}
Monopoles are expected to play an important role in our understanding of confinement of 
quarks in non-Abelian gauge theories. 
To locate these monopole configurations, 't Hooft 
advocated the idea of using zeroes of a composite Higgs \cite{Hoo}. 
In lattice simulations of Yang-Mills theories, one usually looks for monopoles by 
fixing an abelian gauge and looking for U(1) monopoles.
Some popular gauge choices are the maximal Abelian gauge, the laplacian 
Abelian gauge etc. \cite{MAG,LAG}. Therefore, even though 't Hooft's original
proposal was gauge invariant, on the lattice one usually looks for a gauge dependent
quantity. This is however not very satisfactory because the number of monopoles
in different gauges do not agree and it is not clear how to relate the numbers 
obtained in the different gauges. 
Therefore, it is important to have a gauge invariant 
way to detect monopoles.

Recently, progress has been made in this direction by Anishetty, Majumdar and 
Sharatchandra \cite{1}. They have given a criterion for characterizing instanton 
configurations in (euclidean) 3-dimensional Yang-Mills theory in a local and 
manifestly gauge invariant way. This was achieved by reformulating the theory in terms of 
gauge invariant variables closely related to gravity. 
We expect that such a criterion would be useful for locating the monopoles 
in the 3+1--dimensional Yang-Mills theory.

The theory in terms of these new variables however does not tell 
us whether these instanton configurations really occur. 
Since this is an important dynamical
question, we look at it
by simulating 3-dimensional SU(2) gauge theory on the 
lattice. We write down the lattice version of the criterion and check for the 
instanton configurations envisaged in \cite{1}.

The rewriting of the SU(2) gauge theory in terms of new variables is also a duality 
transformation as it neatly separates the 
``spin wave " degrees of freedom from the ``topological" ones. We see this separation 
explicitly from our simulation data.

\section{Method}
The partition function for 3-dimensional Yang-Mills theory is
\begin{equation}\label{part}
Z=\int\;{\cal D}A_i^a(x)\;exp\left ( -\frac{1}{4g^2}\int\;d^3 x
F_{ij}^a(x)F_{ij}^a(x)\right ) 
\end{equation}
where 
\begin{equation}
F_{ij}^a=\partial_iA_j^a-\partial_jA_i^a+\epsilon^{abc}A_i^bA_j^c
\end{equation} 
is the usual field strength and $g$ the coupling constant.

To identify the instantons of the theory, one can use the 
orthogonal set of eigenfunctions of a positive symmetric matrix.
Consider the following eigenvalue equation\footnote{Here, $A$ is just a label 
for the eigenvalues and eigenvectors and is not summed over.}
\begin{equation}
I_{ik}(x)\chi_k^A(x)=\lambda^A(x)\chi_i^A(x).
\end{equation}
Here, $I_{ik}(x)$ is defined at each space-time point $x$ by
\begin{equation}\label{def}
I_{ik}(x)=F_{ij}^a(x)F_{kj}^a(x).
\end{equation}
Isolated points where $I_{ik}$ have triply degenerate eigenvalues are special and 
have topological significance. At such points, the eigenvector fields are singular. The 
index of the singular point is the instanton number.
Thus the instantons in any Yang-Mills configuration $A_i^a(x)$ can be located in 
terms of the eigenvectors $\chi_i^A(x)$.
One can also construct coordinate systems using integral curves of $\chi^A(x)$. 
The coordinate singularities of this coordinate system then correspond to the instantons.

For a generic instanton, the eigenvectors have the behaviour shown in figure 
1(a). To locate them, 
we need to look at the behaviour of the eigenvectors over many lattice
spacings. In this paper, we however consider only ``spherical''
instantons as in figure 1(b). 
For the location of those instantons,
it is, on the other hand,
sufficient to look at the behaviour of the eigenvectors
around only one lattice point.
Therefore, it 
is easier to identify them even though they are rather rare. 


In our simulation, we choose the usual Wilson action for SU(2) gauge theory
\begin{equation}
 S_W=\frac{\beta}{2} \sum_{\Box} tr (U_{ij}U_{jk}U_{kl}U_{li})
\end{equation}
where $U_{ij}$ are the link variables and the products of the links are taken 
around plaquettes. Here, the relation between 
$\beta$ and the gauge coupling $g$ in Eq.~(\ref{part})
is $\beta=4/g^2$. 
We measure the elementary plaquette at every site. The plaquette variable 
can be written as 
\begin{equation}
exp\,i\,F_{ij}^a\sigma^a = cos (|F_{ij}|) + i{\hat F}_{ij}^a\sigma^a\, sin(|F_{ij}|).
\end{equation}
Therefore 
\begin{equation}
F_{ij}^a={\hat F}_{ij}^a\, cos^{-1} cos(|F_{ij}|).
\end{equation}
Once we obtain $F_{ij}^a$ in this way, 
we can construct $I_{ij}$ using definition (\ref{def}).

According to the criterion in \cite{1}, at the location of the instanton all three 
eigenvalues of $I_{ij}$ should be 
degenerate and around that point one of the eigenvectors should have a radial behaviour.
On the lattice we do not expect the eigenvalues to become exactly degenerate, but we 
look for sites where the difference between the eigenvalues are less than some
cut-off. 
In continuum, slightly away from the centre of an instanton with spherical symmetry, one of 
the eigenvalues of $I_{ij}$ will become non-degenerate with the other two which would 
still be 
degenerate. The eigenvector corresponding to this eigenvalue will show a radial 
behaviour. On the lattice we look for spherical insatntons by choosing the
eigenvector corresponding to the largest eigenvalue and plotting it at the site 
of the monopole and it's nearest neighbors to check for this radial behaviour.

\section{Results}
In our simulation, we looked at lattice sizes of length between 16 and 64 lattice 
sites and $\beta$ between 1.5 and 6. In three 
dimensions, $\beta$ has dimension of length to leading order. 
Therefore we kept
the ratio between $\beta $ and the lattice size fixed. 
To check our configurations, we looked at the string tension and matched it with the
values quoted in \cite{Tep}.  We also looked at the minimum of the eigenvalues for 
the various lattice sizes. The results are plotted in fig.2. 


From weak coupling expansion, the average plaquette goes as 
\begin{equation}
P=\frac{n_g}{4\beta} + O(\frac{1}{\beta^2})
\end{equation}
where $n_g$ is the number of group generators.
For SU(2), the eigenvalues which are proportional to $P^2$ should 
therefore behave like
\begin{equation}
P^2=\frac{9}{16\beta^2} + O(\frac{1}{\beta^4}).
\end{equation} 
This idicates that our data matches reasonably well with first order
weak coupling expansion. The deviation that we see is due to 
terms of higher order in $1/\beta$ which we neglect. As expected, the fit is 
much better for higher values of $\beta$.  

Equilibriation time increases with lattice size. The largest lattice we considered is 
$64^3$. For this lattice size
the equilibriation time is of the order of $240$ updates. Therefore we ignored the first 
$300$ updates and after that took measurements in every
successive update for the next 300 updates. We checked that
for this particular measurement autocorrelation effects are negligible.

To choose our cutoff for the difference between the eigenvalues, we looked at the
distribution of the smallest eigenvalues for the various lattice sizes. Then we chose
the cut-off to be half of the mean minimum eigenvalue for each lattice size. We found
that the eigenvalue criterion is extremely sensitive to the choice of cut-off. The
various values for which we took our data are shown in table 1. The first and the
second columns show the various values of $\beta$ and the lttice size
respectively. We recorded the lowest eigenvalue of $I_{ij}$ on the lattice for
every measurement averaged it over the 300 measurements that we took. That is
tabulated in the third column. The fourth column is the cut-off we chose for each
lattice size. It's values are half of that of the third.

With these parameters, we typically found one
lattice site in every two or three measurements which had the difference of
eigenvalues less than the cut-off.  For a few configurations (roughly 3 or 4 out of
the 300 configurations probed), for every lattice size, we found two or three sites in a 
single configuration which met the eigenvalue criterion.  After this we looked at the
eigenvectors corresponding to the largest eigenvalue around the lattice site which
sa\-tis\-fied the criterion for the degenerate eigenvalues. 
For a few of them we found that
there are at least three non-coplanar eigenvectors which converge to or diverge
from a point. Fig.~3 shows 
an eigenvector configuration where three eigenvectors converge 
to the point $C$. To find out whether this is a true congruence in three dimensions, we
rotated the configuration and checked if the eigenvectors remained convergent at various 
different angles. Our data is presented in table 2. Here again the first two
columns show the values of $\beta$ and lattice size. The total number of cases
when the difference between the eigenvalues were less than the cut-off, over the
300 measurements, is tabulated in the third column. Among these configurations, we
looked for radial behaviour of the eigenvector corresponding to the largest
eigenvalue. That number is tabulated in column four.


In figure 4 we look at the distribution of the eigenvectors in the x-y plane. A close 
look shows clearly that most of the spins suffer small deviations and are like spin 
wave excitations. 
However, there are a few spin configurations which form vortices. 
In the figure the two closed boxes show vortex configurations over one plaquette
and two plaquettes respectively. 
We see that the number of vortices extended over more than one plaquette
are more common than those over a single plaquette. We expect similar conclusion to 
hold true for the monopoles as well. This suggests that it is important to try and 
characterize big monopoles as well. That will be investigated later. 


\section{Conclusions}

In this note we have looked at simulation data for SU(2) lattice gauge theory in three 
dimensions. We found that there indeed exist configurations that correspond to the 
criterion constructed in \cite{1}. The data also shows that large size objects extending 
over many cells or plaquettes are favoured compared to the ones 
over single cell or plaquette. 
Also only about $\frac{1}{3}$ of the instantons have a spherical symmetry.

\section{Acknowledgements}

One of the authors (P.M.) would like to thank Profs. Ramesh Anishetty, N.D. Hari Dass and
H.S.Sharatchandra for improvement of the text and several useful suggestions.

\onecolumn

\begin{table}{Table-1}
\begin{tabular}{cccc}
$\beta$ & lattice size & mean lowest & cut-off \\
&&eigenvalue& \\
1.5 & 16 & 0.253 & 0.1265 \\
2.25 & 24 & 0.129 & 0.0646 \\
3 & 32 & 0.079 & 0.0397 \\
3.75 & 40 & 0.0521 & 0.026 \\
4.5 & 48 & 0.0368 & 0.0184 \\
5.25 & 56 & 0.0264 & 0.0132 \\
6 & 64 & 0.0217 & 0.0109 \\
\end{tabular}
\end{table}

\begin{table}{Table-2}   
\begin{tabular}{cccc}
$\beta$ & lattice size & satisfies & satisfies \\
&&eigenvalue&eigenvector \\
1.5 & 16 & 59 & 22 \\
2.25 & 24 & 58 & 12 \\
3& 32 & 67 & 6 \\
3.75 & 40 & 63 & 14 \\
4.5 & 48 & 78 & 16 \\
5.25 & 56 & 52 & 7 \\
6 & 64 & 69 & 18
\end{tabular}
\end{table}

\myfigure{\epsfysize 2.5in\epsfbox{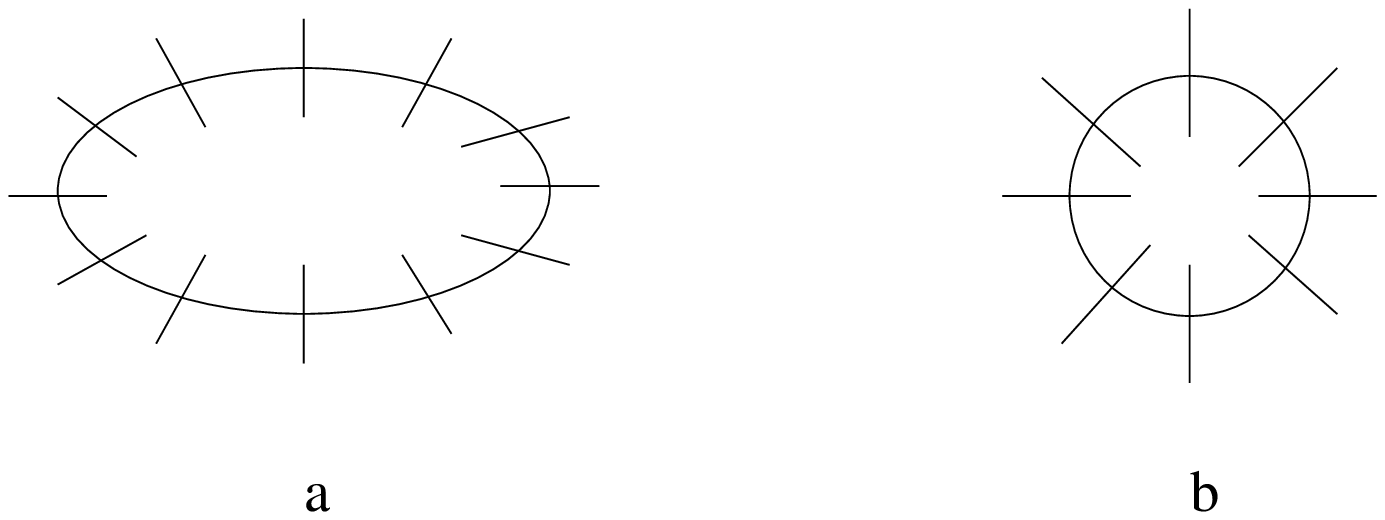}}{}

fig. 1. (a) Eigenvectors around an instanton extended over many lattice spacings.\\
(b). Behavior of the eigenvectors around a ``spherical" instanton centered on one
lattice point.

\newpage
\setlength{\unitlength}{0.240900pt}
\ifx\plotpoint\undefined\newsavebox{\plotpoint}\fi
\sbox{\plotpoint}{\rule[-0.200pt]{0.400pt}{0.400pt}}%
\begin{picture}(1500,900)(0,0)
\font\gnuplot=cmr10 at 10pt
\gnuplot
\put(140.0,123.0){\rule[-0.200pt]{4.818pt}{0.400pt}}
\put(120,123){\makebox(0,0)[r]{0}}
\put(1419.0,123.0){\rule[-0.200pt]{4.818pt}{0.400pt}}
\put(140.0,246.0){\rule[-0.200pt]{4.818pt}{0.400pt}}
\put(120,246){\makebox(0,0)[r]{0.05}}
\put(1419.0,246.0){\rule[-0.200pt]{4.818pt}{0.400pt}}
\put(140.0,369.0){\rule[-0.200pt]{4.818pt}{0.400pt}}
\put(120,369){\makebox(0,0)[r]{0.1}}
\put(1419.0,369.0){\rule[-0.200pt]{4.818pt}{0.400pt}}
\put(140.0,492.0){\rule[-0.200pt]{4.818pt}{0.400pt}}
\put(120,492){\makebox(0,0)[r]{0.15}}
\put(1419.0,492.0){\rule[-0.200pt]{4.818pt}{0.400pt}}
\put(140.0,614.0){\rule[-0.200pt]{4.818pt}{0.400pt}}
\put(120,614){\makebox(0,0)[r]{0.2}}
\put(1419.0,614.0){\rule[-0.200pt]{4.818pt}{0.400pt}}
\put(140.0,737.0){\rule[-0.200pt]{4.818pt}{0.400pt}}
\put(120,737){\makebox(0,0)[r]{0.25}}
\put(1419.0,737.0){\rule[-0.200pt]{4.818pt}{0.400pt}}
\put(140.0,860.0){\rule[-0.200pt]{4.818pt}{0.400pt}}
\put(120,860){\makebox(0,0)[r]{0.3}}
\put(1419.0,860.0){\rule[-0.200pt]{4.818pt}{0.400pt}}
\put(140.0,123.0){\rule[-0.200pt]{0.400pt}{4.818pt}}
\put(140,82){\makebox(0,0){1.5}}
\put(140.0,840.0){\rule[-0.200pt]{0.400pt}{4.818pt}}
\put(284.0,123.0){\rule[-0.200pt]{0.400pt}{4.818pt}}
\put(284,82){\makebox(0,0){2}}
\put(284.0,840.0){\rule[-0.200pt]{0.400pt}{4.818pt}}
\put(429.0,123.0){\rule[-0.200pt]{0.400pt}{4.818pt}}
\put(429,82){\makebox(0,0){2.5}}
\put(429.0,840.0){\rule[-0.200pt]{0.400pt}{4.818pt}}
\put(573.0,123.0){\rule[-0.200pt]{0.400pt}{4.818pt}}
\put(573,82){\makebox(0,0){3}}
\put(573.0,840.0){\rule[-0.200pt]{0.400pt}{4.818pt}}
\put(717.0,123.0){\rule[-0.200pt]{0.400pt}{4.818pt}}
\put(717,82){\makebox(0,0){3.5}}
\put(717.0,840.0){\rule[-0.200pt]{0.400pt}{4.818pt}}
\put(862.0,123.0){\rule[-0.200pt]{0.400pt}{4.818pt}}
\put(862,82){\makebox(0,0){4}}
\put(862.0,840.0){\rule[-0.200pt]{0.400pt}{4.818pt}}
\put(1006.0,123.0){\rule[-0.200pt]{0.400pt}{4.818pt}}
\put(1006,82){\makebox(0,0){4.5}}
\put(1006.0,840.0){\rule[-0.200pt]{0.400pt}{4.818pt}}
\put(1150.0,123.0){\rule[-0.200pt]{0.400pt}{4.818pt}}
\put(1150,82){\makebox(0,0){5}}
\put(1150.0,840.0){\rule[-0.200pt]{0.400pt}{4.818pt}}
\put(1295.0,123.0){\rule[-0.200pt]{0.400pt}{4.818pt}}
\put(1295,82){\makebox(0,0){5.5}}
\put(1295.0,840.0){\rule[-0.200pt]{0.400pt}{4.818pt}}
\put(1439.0,123.0){\rule[-0.200pt]{0.400pt}{4.818pt}}
\put(1439,82){\makebox(0,0){6}}
\put(1439.0,840.0){\rule[-0.200pt]{0.400pt}{4.818pt}}
\put(140.0,123.0){\rule[-0.200pt]{312.929pt}{0.400pt}}
\put(1439.0,123.0){\rule[-0.200pt]{0.400pt}{177.543pt}}
\put(140.0,860.0){\rule[-0.200pt]{312.929pt}{0.400pt}}
\put(789,21){\makebox(0,0){$\beta$}}
\put(140.0,123.0){\rule[-0.200pt]{0.400pt}{177.543pt}}
\put(1279,820){\makebox(0,0)[r]{Lowest eigenvalue}}
\put(1299.0,820.0){\rule[-0.200pt]{24.090pt}{0.400pt}}
\put(1299.0,810.0){\rule[-0.200pt]{0.400pt}{4.818pt}}
\put(1399.0,810.0){\rule[-0.200pt]{0.400pt}{4.818pt}}
\put(140.0,725.0){\rule[-0.200pt]{0.400pt}{9.395pt}}
\put(130.0,725.0){\rule[-0.200pt]{4.818pt}{0.400pt}}
\put(130.0,764.0){\rule[-0.200pt]{4.818pt}{0.400pt}}
\put(357.0,431.0){\rule[-0.200pt]{0.400pt}{4.577pt}}
\put(347.0,431.0){\rule[-0.200pt]{4.818pt}{0.400pt}}
\put(347.0,450.0){\rule[-0.200pt]{4.818pt}{0.400pt}}
\put(573.0,312.0){\rule[-0.200pt]{0.400pt}{2.891pt}}
\put(563.0,312.0){\rule[-0.200pt]{4.818pt}{0.400pt}}
\put(563.0,324.0){\rule[-0.200pt]{4.818pt}{0.400pt}}
\put(790.0,247.0){\rule[-0.200pt]{0.400pt}{1.927pt}}
\put(780.0,247.0){\rule[-0.200pt]{4.818pt}{0.400pt}}
\put(780.0,255.0){\rule[-0.200pt]{4.818pt}{0.400pt}}
\put(1006.0,210.0){\rule[-0.200pt]{0.400pt}{1.445pt}}
\put(996.0,210.0){\rule[-0.200pt]{4.818pt}{0.400pt}}
\put(996.0,216.0){\rule[-0.200pt]{4.818pt}{0.400pt}}
\put(1223.0,186.0){\rule[-0.200pt]{0.400pt}{0.964pt}}
\put(1213.0,186.0){\rule[-0.200pt]{4.818pt}{0.400pt}}
\put(1213.0,190.0){\rule[-0.200pt]{4.818pt}{0.400pt}}
\put(1439.0,175.0){\rule[-0.200pt]{0.400pt}{0.723pt}}
\put(1429.0,175.0){\rule[-0.200pt]{4.818pt}{0.400pt}}
\put(140,745){\raisebox{-.8pt}{\makebox(0,0){$\Diamond$}}}
\put(357,440){\raisebox{-.8pt}{\makebox(0,0){$\Diamond$}}}
\put(573,318){\raisebox{-.8pt}{\makebox(0,0){$\Diamond$}}}
\put(790,251){\raisebox{-.8pt}{\makebox(0,0){$\Diamond$}}}
\put(1006,213){\raisebox{-.8pt}{\makebox(0,0){$\Diamond$}}}
\put(1223,188){\raisebox{-.8pt}{\makebox(0,0){$\Diamond$}}}
\put(1439,176){\raisebox{-.8pt}{\makebox(0,0){$\Diamond$}}}
\put(1349,820){\raisebox{-.8pt}{\makebox(0,0){$\Diamond$}}}
\put(1429.0,178.0){\rule[-0.200pt]{4.818pt}{0.400pt}}
\put(1279,779){\makebox(0,0)[r]{Perturbation theory}}
\multiput(1299,779)(20.756,0.000){5}{\usebox{\plotpoint}}
\put(1399,779){\usebox{\plotpoint}}
\put(140,737){\usebox{\plotpoint}}
\multiput(140,737)(7.227,-19.457){2}{\usebox{\plotpoint}}
\multiput(153,702)(7.607,-19.311){2}{\usebox{\plotpoint}}
\multiput(166,669)(8.253,-19.044){2}{\usebox{\plotpoint}}
\put(187.26,621.85){\usebox{\plotpoint}}
\put(196.66,603.35){\usebox{\plotpoint}}
\multiput(206,586)(10.213,-18.069){2}{\usebox{\plotpoint}}
\put(227.21,549.11){\usebox{\plotpoint}}
\put(238.18,531.49){\usebox{\plotpoint}}
\put(249.65,514.20){\usebox{\plotpoint}}
\put(261.63,497.25){\usebox{\plotpoint}}
\put(274.24,480.77){\usebox{\plotpoint}}
\put(287.07,464.46){\usebox{\plotpoint}}
\put(300.95,449.05){\usebox{\plotpoint}}
\put(315.63,434.37){\usebox{\plotpoint}}
\put(330.31,419.69){\usebox{\plotpoint}}
\put(345.62,405.71){\usebox{\plotpoint}}
\put(361.46,392.30){\usebox{\plotpoint}}
\put(377.36,378.96){\usebox{\plotpoint}}
\put(393.99,366.55){\usebox{\plotpoint}}
\put(411.26,355.05){\usebox{\plotpoint}}
\put(428.88,344.07){\usebox{\plotpoint}}
\put(446.71,333.46){\usebox{\plotpoint}}
\put(464.98,323.62){\usebox{\plotpoint}}
\put(483.33,313.92){\usebox{\plotpoint}}
\put(502.17,305.23){\usebox{\plotpoint}}
\put(521.20,296.92){\usebox{\plotpoint}}
\put(540.39,289.05){\usebox{\plotpoint}}
\put(559.58,281.16){\usebox{\plotpoint}}
\put(579.25,274.59){\usebox{\plotpoint}}
\put(598.93,268.02){\usebox{\plotpoint}}
\put(618.45,261.01){\usebox{\plotpoint}}
\put(638.60,256.09){\usebox{\plotpoint}}
\put(658.57,250.48){\usebox{\plotpoint}}
\put(678.54,244.88){\usebox{\plotpoint}}
\put(698.77,240.21){\usebox{\plotpoint}}
\put(718.99,235.54){\usebox{\plotpoint}}
\put(739.25,231.02){\usebox{\plotpoint}}
\put(759.67,227.38){\usebox{\plotpoint}}
\put(779.89,222.72){\usebox{\plotpoint}}
\put(800.36,219.33){\usebox{\plotpoint}}
\put(820.88,216.17){\usebox{\plotpoint}}
\put(841.22,212.11){\usebox{\plotpoint}}
\put(861.74,209.04){\usebox{\plotpoint}}
\put(882.32,206.44){\usebox{\plotpoint}}
\put(902.88,203.71){\usebox{\plotpoint}}
\put(923.40,200.55){\usebox{\plotpoint}}
\put(944.03,198.42){\usebox{\plotpoint}}
\put(964.65,196.18){\usebox{\plotpoint}}
\put(985.23,193.60){\usebox{\plotpoint}}
\put(1005.82,191.03){\usebox{\plotpoint}}
\put(1026.51,189.42){\usebox{\plotpoint}}
\put(1047.09,186.84){\usebox{\plotpoint}}
\put(1067.79,185.30){\usebox{\plotpoint}}
\put(1088.48,183.73){\usebox{\plotpoint}}
\put(1109.08,181.30){\usebox{\plotpoint}}
\put(1129.76,179.56){\usebox{\plotpoint}}
\put(1150.45,177.97){\usebox{\plotpoint}}
\put(1171.15,176.42){\usebox{\plotpoint}}
\put(1191.85,174.86){\usebox{\plotpoint}}
\put(1212.57,174.00){\usebox{\plotpoint}}
\put(1233.27,172.67){\usebox{\plotpoint}}
\put(1253.97,171.08){\usebox{\plotpoint}}
\put(1274.66,169.52){\usebox{\plotpoint}}
\put(1295.40,168.97){\usebox{\plotpoint}}
\put(1316.09,167.38){\usebox{\plotpoint}}
\put(1336.80,166.00){\usebox{\plotpoint}}
\put(1357.52,165.19){\usebox{\plotpoint}}
\put(1378.23,164.00){\usebox{\plotpoint}}
\put(1398.95,163.08){\usebox{\plotpoint}}
\put(1419.68,162.49){\usebox{\plotpoint}}
\put(1439,161){\usebox{\plotpoint}}
\end{picture}

fig. 2. In this figure we plot
the lowest eigenvalues of $I_{ij}$ (points with errorbars) along the y-axis for
various values of $\beta$ (along the x-axis).
The dashed line is first order weak
coupling expansion for the square of the plaquette $P^2=\frac{9}{16 \beta^2}$.

\newpage
\myfigure{\epsfysize 2.6in\epsfbox{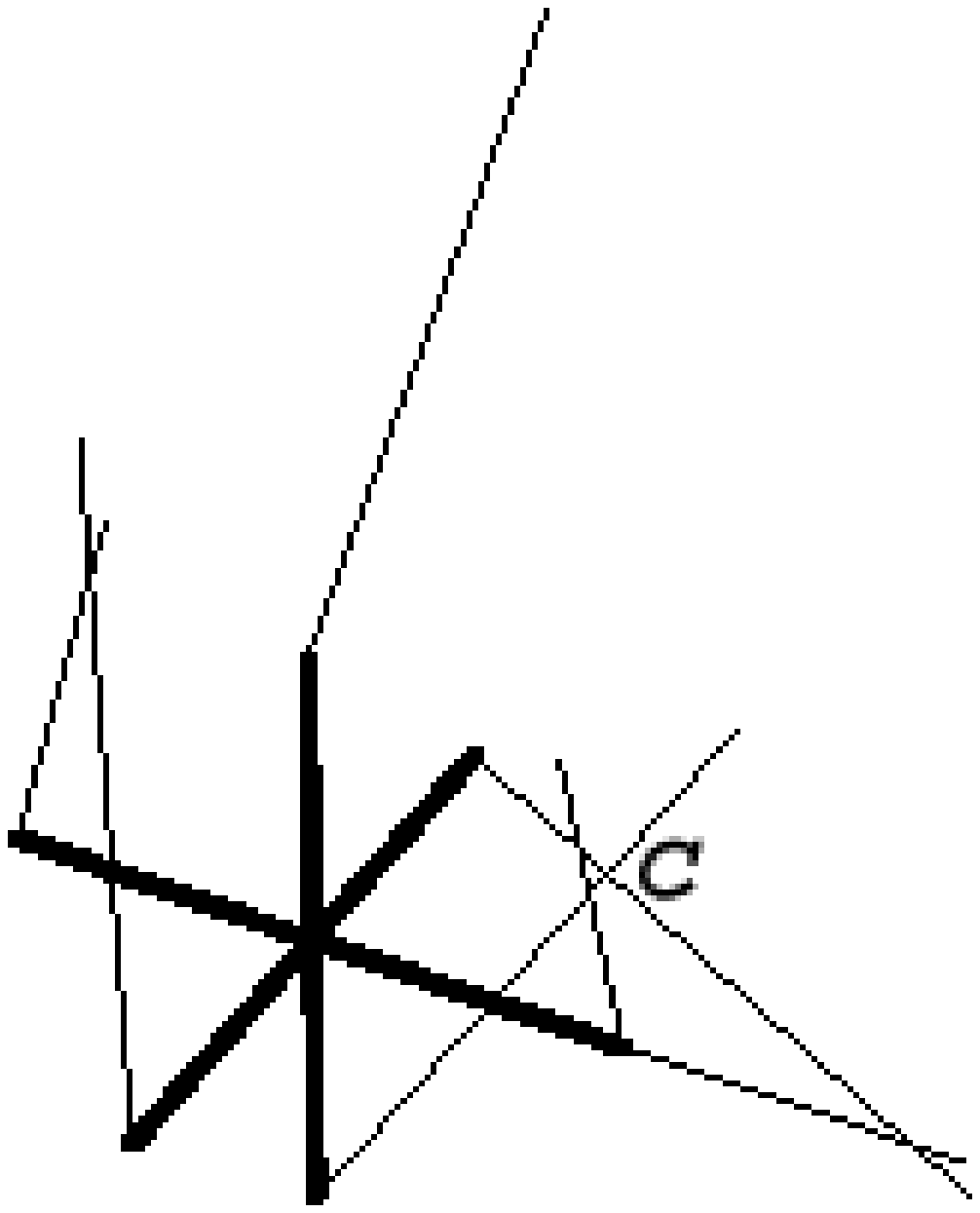}}{}

fig. 3. Eigenvector configuration
we are looking for. The eigenvectors shown here are scaled to twice their size to
show their crossing explicitly. Lattice size is 56.

\newpage
\myfigure{\epsfysize 6in\epsfbox{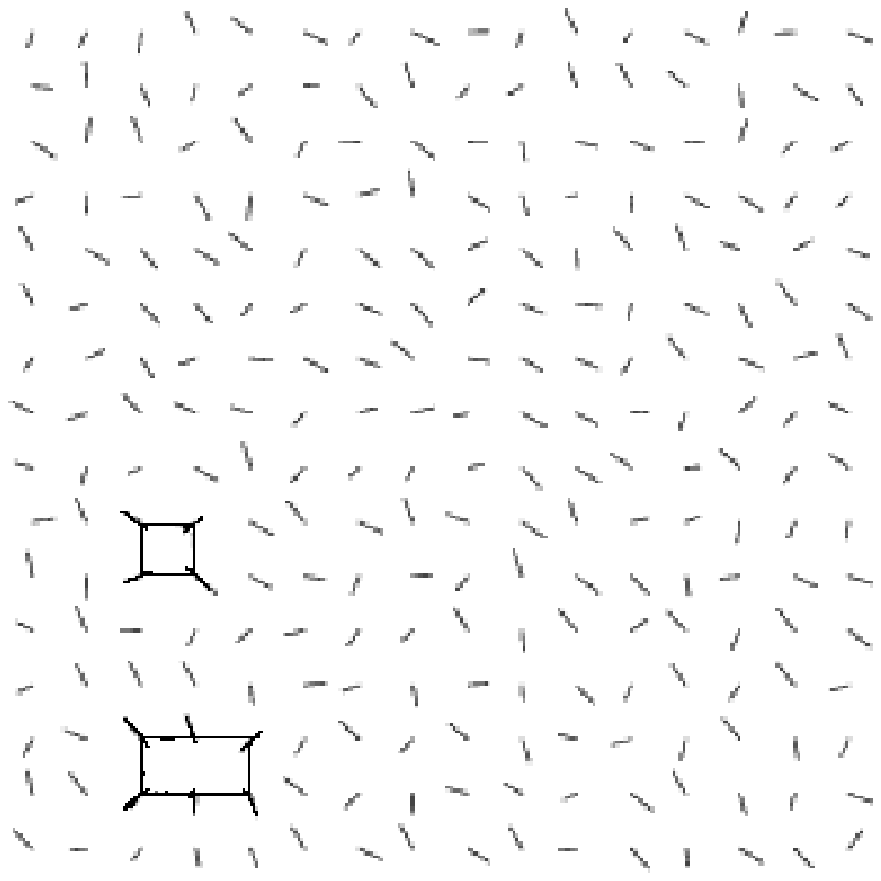}}{}

fig. 4. Snapshot of a x-y plane with 
eigenvectors projected to the plane. Closed boxes show projection of spherical
instanton formed on one plaquette and ``non-spherical" instanton extended over 
two plaquettes on the plane. Lattice size is 16.

\end{document}